\documentclass[
 reprint,
longbibliography ,
 aps,
  twoside,
 superscriptaddress,
bibnotes,
pra
]{revtex4-1}

\usepackage{filecontents}

\usepackage{physics}
\usepackage{dsfont} 
\usepackage{color,newfloat}
\usepackage{graphicx}
\usepackage{dcolumn}
\usepackage{bm}
\usepackage{amsmath,amssymb}

\usepackage{chngcntr} 

\usepackage{array}  
\usepackage{hhline}
\usepackage{multirow}
\usepackage{afterpage}
\usepackage[normalem]{ulem}

\usepackage[table,xcdraw]{xcolor}

\def\({\left(}
\def\){\right)}
\def\[{\left[}
\def\]{\right]}

\begin{document}


\title{
Near-ideal spontaneous photon sources in silicon quantum photonics
}

\author{S. Paesani}
\altaffiliation{Contributed equally.}
\affiliation{Quantum Engineering Technology Labs, H. H. Wills Physics Laboratory and Department of Electrical and Electronic Engineering, University of Bristol, Bristol BS81FD, UK}

\author{M. Borghi}
\altaffiliation{Contributed equally.}
\affiliation{Quantum Engineering Technology Labs, H. H. Wills Physics Laboratory and Department of Electrical and Electronic Engineering, University of Bristol, Bristol BS81FD, UK}
\affiliation{SM Optics s.r.l., Research Programs, Via John Fitzgerald Kennedy 2, 20871 Vimercate, Italy}

\author{S. Signorini}
\altaffiliation{Contributed equally.}
\affiliation{Department of Physics, University of Trento, Via Sommarive 14, 38123 Trento, Italy}

\author{A. Ma\"inos}
\affiliation{Quantum Engineering Technology Labs, H. H. Wills Physics Laboratory and Department of Electrical and Electronic Engineering, University of Bristol, Bristol BS81FD, UK}

\author{L. Pavesi}
\affiliation{Department of Physics, University of Trento, Via Sommarive 14, 38123 Trento, Italy}

\author{A. Laing}
\affiliation{Quantum Engineering Technology Labs, H. H. Wills Physics Laboratory and Department of Electrical and Electronic Engineering, University of Bristol, Bristol BS81FD, UK}


\begin{abstract}
While integrated photonics is
a robust platform
for quantum information processing,
architectures for photonic quantum computing
place stringent demands on high quality information carriers.
Sources of single photons
that are 
highly indistinguishable and pure,
that are either near-deterministic or heralded with high efficiency,
and that are suitable for
mass-manufacture,
have been elusive.
Here, we demonstrate on-chip photon sources that
simultaneously meet each of these requirements.
Our photon sources are fabricated
in silicon
using mature processes,
and exploit a novel
dual-mode pump-delayed excitation scheme
to engineer 
the emission of spectrally pure photon pairs through intermodal spontaneous four-wave mixing in low-loss spiralled multi-mode waveguides. 
We simultaneously measure 
a spectral purity of
$0.9904 \pm 0.0006$,
a mutual indistinguishably of
$0.987 \pm 0.002$,
and $>90\%$ intrinsic heralding efficiency. 
We measure on-chip quantum interference with a visibility of $0.96 \pm 0.02$ between heralded photons from different sources. 
These results represent a decisive step for scaling quantum information processing in integrated photonics.
\end{abstract}


\maketitle



Sustained progress
in the engineering of platforms for quantum information processing
has recently achieved
a scale that surpasses
the capabilities
of classical computers to solve
specialised and abstract problems
~\cite{preskill2018,Bernien2017,Google,BS_20phot}.
But while achieving a computational advantage
for practical or industrially relevant problems
may be possible with further scaling of 
special purpose NISQ
(noisy intermediate scale quantum)
devices~\cite{preskill2018},
more general purpose quantum computers
will require a hardware platform
that integrates millions of components,
individually operating above some fidelity threshold~\cite{reiher2017,gidney2019}.
Silicon quantum photonics~\cite{silverstone2016silicon},
which is compatible with
complementary metal-oxide-semiconductor (CMOS) fabrication,
provides a potential platform
for very large-scale
quantum information processing~\cite{rudolph2017optimistic,wang2018multidimensional,paesani2019}.

All-photonic quantum computing architectures
rely on arrays of many photon sources
to achieve combinatorial speed-ups in
quantum sampling algorithms~\cite{Lund2014,GBS_Main},
or to approximate
an on-demand source of
single photons~\cite{Pittman2002,Migdall2002,kaneda2019},
and supply entangling circuitry for general purpose
quantum computing~\cite{Gimeno-Segovia2015,rudolph2017optimistic}.
In the former case,
the level of indistinguishability among photons
upper bounds the computational complexity of sampling algorithms \cite{renema2018};
in the latter case,
photon impurity and distinguishability
lead to logical errors~\cite{Gimeno-Segovia2015,SparrowNoise}.
Furthermore,
and in general,
lossy or inefficient heralding of photons
vitiates the scaling of
photonic quantum information processing.
The lack of a demonstration that
simultaneously overcomes all of these challenges
has been a bottleneck to scalability
for quantum computing in integrated photonics.

Progress in solid-state sources of single photons
make quantum dots an attractive approach for certain NISQ experiments~\cite{Somaschi2016,wang2019towards}. 
However, the low-loss integration
of solid-state sources into photonic circuitry,
that maintains distinguishability over many photons,
is an ongoing challenge~\cite{Laucht2012,arcari2014near}.
Integrated sources of photons based on spontaneous processes,
such as four-wave mixing (SFWM)
in single-mode waveguides
or micro-ring cavities~\cite{silverstone2016silicon,caspani2017integrated}
are appealing for their
manufacturability.
However,
spontaneous sources incur
limitations
to purity
and to heralding efficiency~\cite{caspani2017integrated},
with micro-ring cavities
additionally requiring resonance tuning
to avoid distinguishability among different cavities~\cite{carolan2019scalable,Llewellyn2019}.

\begin{figure*}[t!]
  \centering
  \includegraphics[
  trim=0 0 0 -10,
  width=1 \textwidth]{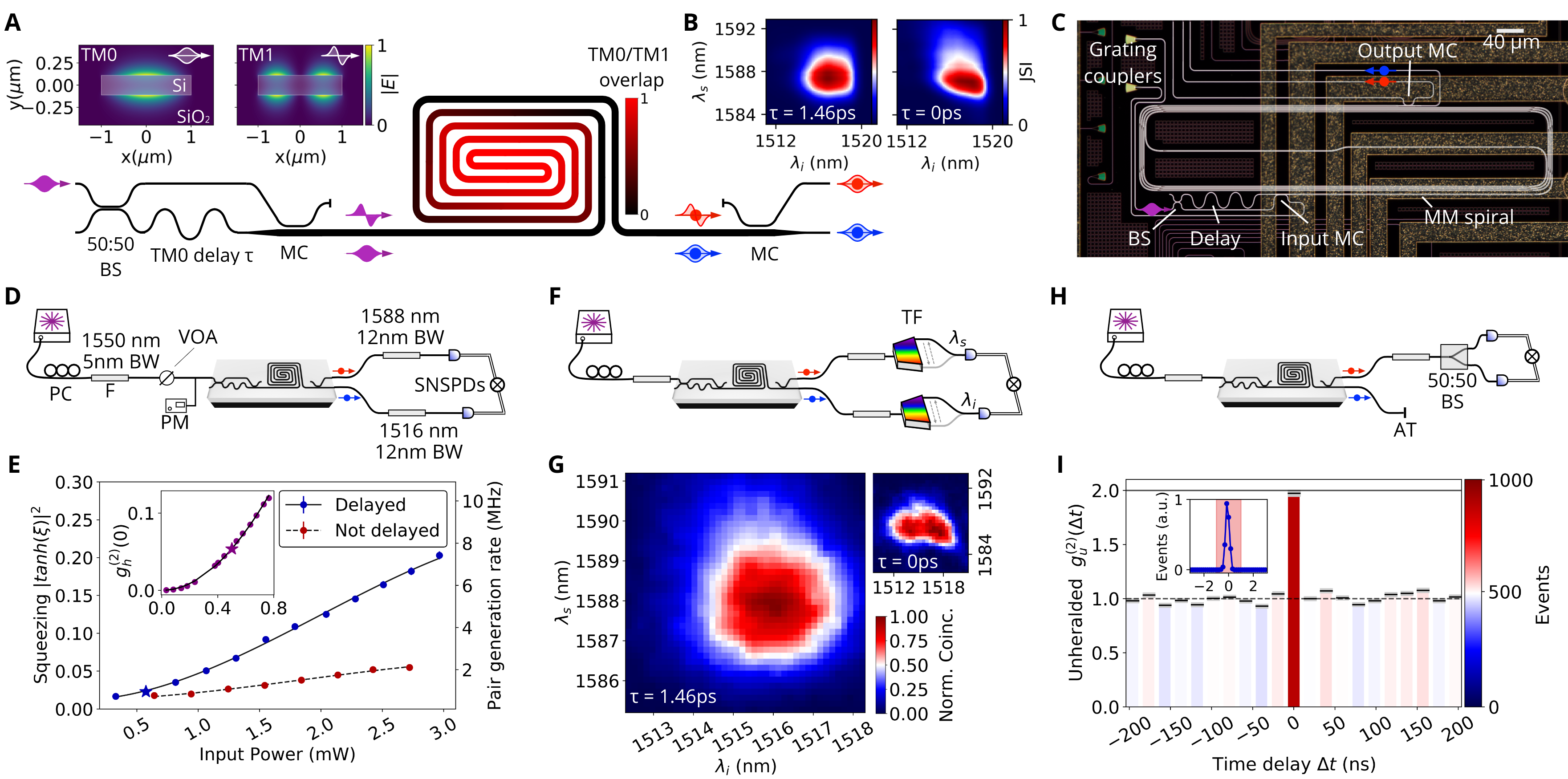}
  \caption{  
  \textbf{Design and performance of the multi-modal source.} 
  \textbf{A.} Schematic of the source. 
  An input near-1550~nm pulsed pump laser ($4.5$~nm bandwidth), initially propagating in the TM0 mode, is split using a 50:50 beam-splitter (BS). 
  The output in the upper arm of the BS is converted into the TM1 mode via a mode-converter (MC), while the TM0 output in the lower arm is delayed by a time $\tau=1.46$~ps.
  Due to the different group velocities, the two modes become overlapped and subsequently diverge again while propagating through the source, as qualitatively colour-coded in the figure. 
  Photon pairs, with the signal photon (near $1588$~nm) in the TM1 mode and the idler photon (near $1516$~nm) in the TM0 mode, are emitted via inter-modal SFWM and finally deterministically separated via a MC. 
  Inset: cross sections of the TM0 and TM1 modes in the MM waveguide. 
   \textbf{B.} Simulated JSI of the source in the presence of a delay $\tau=1.46$~ps (left) and with no delay (right), with corresponding single photon purities of 99.4\% and 84.0\%. 
   \textbf{C.} Optical microscope image of a single multi-modal source structure - waveguides are highlighted. 
   \textbf{D.} Set-up to characterise squeezed light
   via second-order correlation measurements,
   using a polarisation controller (PC),
   fiber pass-band filter (F),
   variable optical attenuator (VOA),
   and an optical power monitor (PM).
   \textbf{E.} Measured squeezing as a function of (off-chip) pump power. 
   Blue and red points are data measured in a source with and without delay, respectively,
   with a fit shown as a black line. 
   The stars indicate the  typical operating regime. 
   Inset: measured heralded $g_h^{(2)}(0)$ as a function of input powers. 
   \textbf{F.} Set-up for the characterisation of the emitted JSI, using a tuneable filter (TF).
   \textbf{G.} Measured JSI from the source with delay (left) and without delay (right), with respective corresponding spectral purities of $0.9904(6)$ and $0.931(2)$.
   \textbf{H.} Set-up for purity characterisation via unheralded second-order correlation measurements. Idler photons are discarded via an absorbing termination (AT).
   \textbf{I.} Measured unheralded $g_u^{(2)}(\Delta t)$ in the source with delay. 
   Each bar corresponds to a coincidence window of $2$~ns (inset). 
   The measured $g_u^{(2)}(0)=1.97(3)$ corresponds to a photon spectral purity of $0.97(3)$.
   Error bars are calculated assuming Poissonian error statistics.
   } 
  \label{Fig_Purity}
\end{figure*} 

\begin{figure*}[t]
  \centering
  \includegraphics[
  trim=0 0 0 -10,
  width=1 \textwidth]{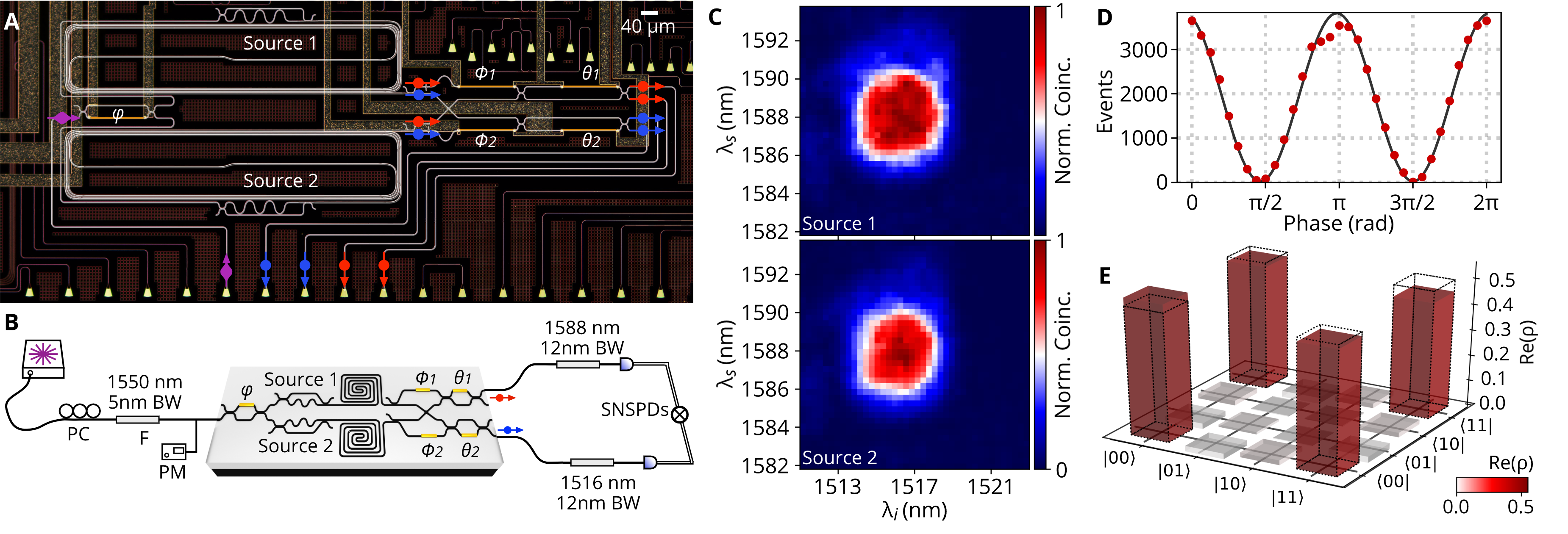}
  \caption{
  \textbf{Multiple sources and indistinguishability characterisation.}
  \textbf{A.} Optical microscope image of the device for the coherent pumping of two sources and processing of the emitted photons. 
  Input pump light is split between the two sources via an input MZI.
  Crossers are used to group the signal and idler photons emitted from both sources, and arbitrary unitary operations on signals (idlers) are implemented via a phase-shifter $\phi_1$ ($\phi_2$) and a MZI with internal phase $\theta_1$ ($\theta_2$). 
  \textbf{B.} Schematic of the integrated circuit and set-up to characterise the indistinguishability between the sources. 
  \textbf{C.} Individual JSIs measured with separate pumping of source 1 (top panel) and source 2 (bottom panel). 
  The indistinguishability of the two measured spectra, calculated with the overlap of the two JSIs, is 0.985(1).
  \textbf{D.} Measured reverse-HOM fringe from the two sources. 
  Error bars, from Poissonian photon statistics, are smaller than markers. 
  The fringe visibility, which corresponds to source indistinguishability, is 0.987(2).
  \textbf{E.} Density matrix of the two-qubit entangled state obtained when coherently pumping the two sources,
  reconstructed via quantum state tomography. 
  Fidelity with the ideal state $\ket{\Phi_+}=(\ket{00}+\ket{11})/\sqrt{2}$, pictured as transparent bars, is $0.989(3)$.
  The source indistinguishability inferred from the tomography is $0.982(6)$.
  } 
  \label{Fig_Indist}
\end{figure*} 

Here, we demonstrate the engineering of
a CMOS-compatible source
of heralded single photons
using silicon photonics,
which simultaneously meets the requirements
for scalable quantum photonics:
high purity, high heralding efficiency, and high indistinguishabilty.
The source,
depicted in Fig.~\href{\ref{Fig_Purity}}{\ref{Fig_Purity}a},
is based on inter-modal SFWM,
where phase-matching is engineered
by propagating the pump in different transverse modes
of a spiralled multi-mode (MM) waveguide~\cite{signorini2018intermodal}.

Integrated photon sources in silicon
are based on SFWM,
where, 
if phase-matching (momentum conservation) 
and energy conservation 
conditions are satisfied, 
light from a pump laser can be converted into pairs of single photons~\cite{caspani2017integrated}.
In standard SFWM in single-mode waveguides, 
near-zero dispersion 
produces broad phase-matching bands around the pump wavelength,
such that the process, dominated by energy conservation conditions,
induces undesired strong spectral anticorrelations between the emitted photons.  
In contrast, in this work we suppress such correlations adopting a novel inter-modal approach to SFWM.
As shown in Fig.~\href{\ref{Fig_Purity}}{\ref{Fig_Purity}a}, 
an input pulsed laser
coherently pumps the two lowest order
transverse magnetic (TM) modes of a MM waveguide,
namely TM0 and TM1
(see Fig.~\href{\ref{Fig_Purity}}{\ref{Fig_Purity}a} inset),
and generates pairs of idler and signal photons in these modes via inter-modal phase-matching.
The dispersion relations between the TM0 and TM1 modes 
are such that a discrete narrow phase-matching band appears~\cite{signorini2018intermodal}.
By tailoring the waveguide cross-section, the modal dispersion can be accurately engineered to design the phase-matching band with a bandwidth similar to the pump bandwidth (related to energy conservation).
This suppresses the frequency anticorrelations imposed by energy conservation, and enhances the spectral purity of the emitted photons~\cite{signorini2018intermodal}.
In particular, we exploit the different modal group velocities in silicon waveguides to achieve a condition where the idler and signal photons are generated on TM0 at $\simeq 1516$~nm and on TM1 at $\simeq 1588$~nm, respectively,
with a bandwidth of approximately $4$~nm (see Supplementary Information~1).

Moreover, to obtain a near-unit spectral purity,
we further suppress residual correlations in the joint spectrum
by inserting a delay $\tau$ on
the TM0 component of the pump (with higher group velocity than TM1)
before injecting it in the source.
The delay gradually increases and decreases
the temporal overlap between the pump
in the TM0 and TM1 modes
along the multi-modal waveguide source
(colour-coded in Fig.~\href{\ref{Fig_Purity}}{\ref{Fig_Purity}a}).
This results in an adiabatic switching
of non-linear interactions in the source,
which suppresses spurious
spectral correlations~\cite{fang2013state,zhang2019dual}.
Simulations
(see Supplementary Information~1)
predict a spectral purity of $99.4\%$
in this configuration,
in contrast to the case where no delay is applied,
which predicts a purity of $84.0\%$,
as shown in Fig.~\href{\ref{Fig_Purity}}{\ref{Fig_Purity}b}.

%
Figure ~\href{\ref{Fig_Purity}}{\ref{Fig_Purity}c},
shows the compact footprint for the MM waveguide source
obtained by adopting a spiral geometry.
The delayed-pump excitation scheme
is implemented in three stages,
as shown in Fig.~\href{\ref{Fig_Purity}}{\ref{Fig_Purity}a}. 
The pump, initially in TM0,
is split by a balanced beam-splitter;
one arm receives a delay of $\tau$ with respect to the other,
then the two arms are recombined using a TM0 to TM1 mode converter,
and injected in the MM waveguide.
Once generated, the signal photon is separated from the idler via a second TM1 to TM0 mode converter.
After processing,   
signal and idler photons are out-coupled to fibres,
where the pump is filtered out via broad-band fibre bragg-gratings,
and single photons are detected
with superconducting-nanowire single photon detectors (SNSPDs).
%

We experimentally characterised
the squeezing value $\xi$
of the generated two-mode squeezed state
emitted from individual sources
with second-order correlation measurements \cite{Christ2011}
(see Supplementary Information~5).
Figure~\href{\ref{Fig_Purity}}{\ref{Fig_Purity}e}
compares
results for both the delayed and the non-delayed cases.
Experimental results confirm our simulations
and additionally demonstrate higher brightness
as a benefit of the temporal delay scheme
(see Supplementary Information~1).
Squeezing values up to
$\left|\tanh(\xi)\right|^2\simeq 0.2$
are observed
using a small input (off-chip)
average pump power of $3$~mW,
corresponding to $>8$~MHz photon-pair generation rates on-chip. 
To reduce noise from multi-photon events,
measurements reported from this point on
are performed with an input pump power of $~0.5$~mW:
coincidence rates are measured at $~15$~kHz,
with heralded single photon $g^{(2)}_h$ measured at $0.053(1)$
(Fig.~\href{\ref{Fig_Purity}}{\ref{Fig_Purity}e} inset).
%

%
Source purity is first estimated from a direct measurement
of the joint spectral intensity (JSI)~\cite{jizan2015bi}. 
The JSI reconstruction is implemented using narrow-bandwidth
tunable filters to scan the emitted wavelengths of the signal and idler photons, as pictured in Fig.~\href{\ref{Fig_Purity}}{\ref{Fig_Purity}f}. 
Data from a source with no temporal delay
yields
a JSI with a spectral purity of $0.931(2)$,
which increases to $99.04(6)\%$,
in the scheme with a delay,
as shown in Fig.~\href{\ref{Fig_Purity}}{\ref{Fig_Purity}g}.
The contrasting measurements show
a clear suppression of spurious correlations
with the delay scheme.
A second estimation of the emitted single photon purity for the delayed structure is performed via unheralded second-order correlation measurements $g_u^{(2)}$~\cite{Christ2011}. 
These are implemented by
dividing the output signal mode with an off-chip
balanced fibre beam-splitter
and measuring coincidences between the two output arms
(see Fig.~\href{\ref{Fig_Purity}}{\ref{Fig_Purity}h}). 
Measured unheralded second order-correlation values are reported in Fig.~\href{\ref{Fig_Purity}}{\ref{Fig_Purity}i}. 
We obtain $g_u^{(2)}(0)= 1.97(3)$, which corresponds to a single photon purity of $97(3)\%$, consistent with the value obtained from the JSI. 
%

The capability of the sources
to generate pure photons with no requirement for filtering
enables the simultaneous achievement
of high heralding efficiency and high purity. 
In our experiment,
off-chip filters are used solely for pump rejection:
their bandwidth ($12$~nm, flat transmission) contains $>99\%$ of the emitted spectra,
which results in ultra-high filtering heralding efficiency~\cite{MeyerScott2017}. 
While the effect of filtering is thus negligible,
the intrinsic heralding efficiency of the source
is affected by linear and non-linear transmission losses
inside the waveguide. 
These losses are, however,
greatly mitigated in MM waveguides
(which present $<-0.5$ dB/cm linear loss,
see Supplementary Information~2).
Taking into account the characterised losses,
we estimate a heralding efficiency of approximately $95\%$
for an individual source. 
The measured heralding efficiency
at the off-chip detectors is $12.6(2)\%$,
corresponding to $91(9)\%$ on-chip intrinsic heralding efficiency
after correcting for the characterised losses
in the channel to the detectors
(see Supplementary Information~2),
which can be highly suppressed
by implementing low-loss off-chip couplers~\cite{Ding2014}
or with integrated detectors~\cite{khasminskaya2016}.

\begin{figure}[t!]
  \centering
  \includegraphics[
  trim=0 0 0 -10,
  width=0.45 \textwidth]{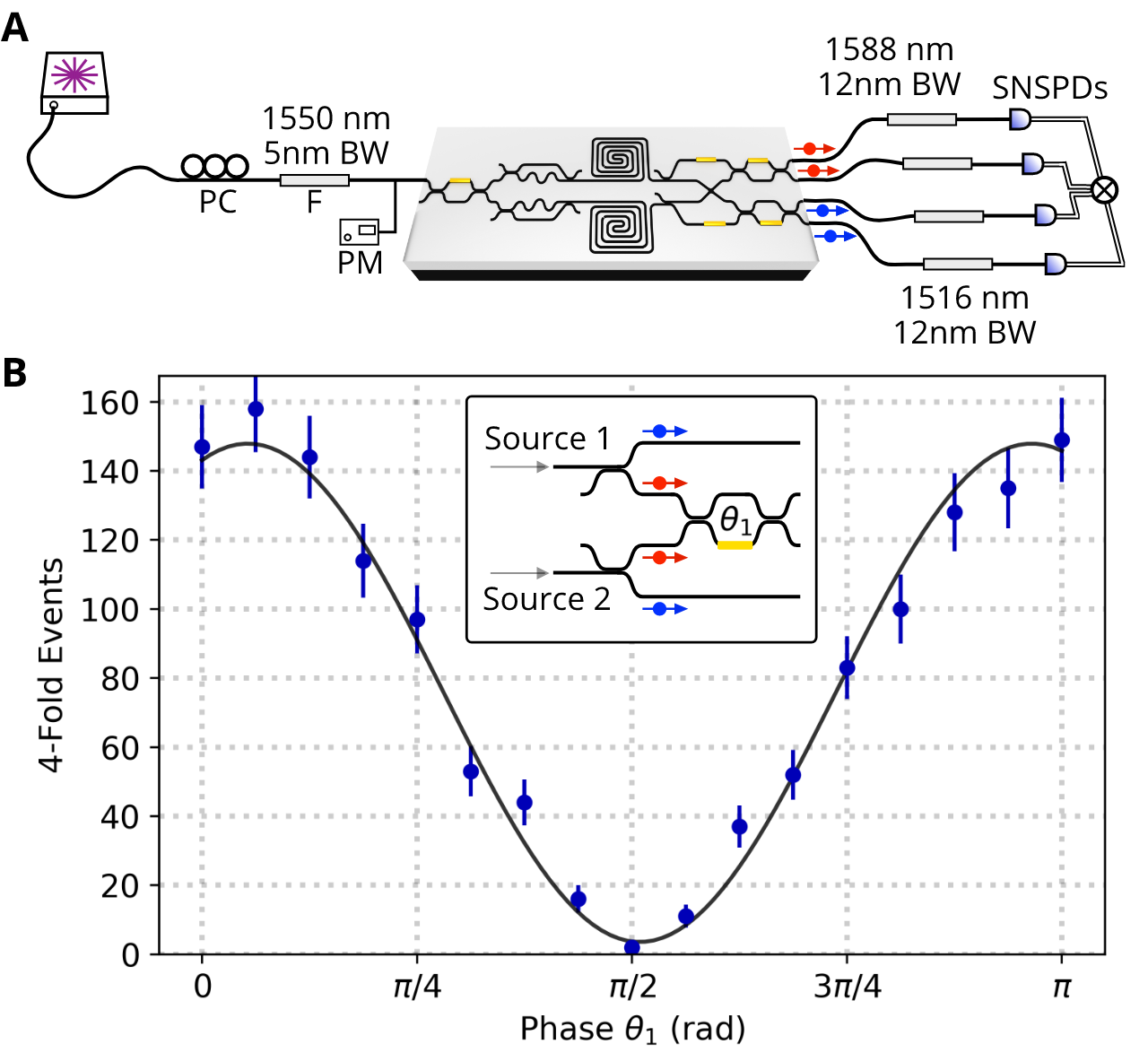}
  \caption{
  \textbf{Heralded Hong-Ou-Mandel interference.}
  \textbf{A.} Experimental setup for 4-photon heralded HOM experiments.
  \textbf{B.} Heralded HOM results. 
  Points are raw four-photon counts measured over 4 hours of integration time each for different values of the MZI phase $\theta$, with a solid line fit to the data, presenting a visibility of $0.96(2)$.
  Error bars consider Poissonian photon statistics.
  Inset: schematic of the integrated circuit configuration
  for measuring the heralded HOM fringe.
  } 
  \label{Fig_HOM}
\end{figure}


To experimentally test the source indistinguishabilty
we integrate a reconfigurable photonic circuit
to perform quantum interference between different sources. 
Schematics of the circuit are shown in Fig.~\href{\ref{Fig_Indist}}{\ref{Fig_Indist}a-b}.
Two sources are coherently pumped by
splitting the input laser
with an on-chip 
tunable
Mach-Zehnder interferometer (MZI);
the resulting idler and signal modes from the different sources
are grouped and interfered on-chip
using additional integrated phase-shifters and MZIs (see Methods). 
Using this circuit,
we experimentally estimate
the indistinguishability among the sources
using three different types of measurements. 
First, we reconstruct the JSI of each source
by operating the two sources individually. 
The overlap of the JSIs reconstructed from each source (Fig.~\href{\ref{Fig_Indist}}{\ref{Fig_Indist}c})
estimates a mutual indistinguishability of $98.5(1)\%$. 

A second measurement of the indistinguishability
was performed via reversed Hong-Ou-Mandel (HOM)
interference between the
two sources~\cite{silverstone2014,silverstone2015,wang2018multidimensional}. 
Both sources were pumped and the respective
idler and signal modes were interfered by
tuning the output MZIs to act as 50:50 beam-splitters. 
The $98.7(2)\%$ visibility of the reversed HOM fringe,
shown in Fig.~\href{\ref{Fig_Indist}}{\ref{Fig_Indist}d}
and obtained by scanning the phases
$\phi_1=\phi_2=\phi$,
corresponds directly to the source indistinguishability
(see Supplementary Information~6). 

A further estimate of indistinguishability is obtained
by testing the entanglement generated when
coherently pumping the two sources~\cite{silverstone2015,wang2018multidimensional}. 
Using quantum state tomography,
we experimentally reconstruct the density matrix shown in Fig.~\href{\ref{Fig_Indist}}{\ref{Fig_Indist}e},
which has a fidelity of $98.9(3)\%$
with the ideal maximally-entangled state
$\ket{\Phi_+}=(\ket{00}+\ket{11})/\sqrt{2}$,
and provides an indistinguishability value of $98.2(6)\%$
(see Supplementary Information~6 for details). 
%

%
A key figure of merit for multi-photon experiments,
particularly in the context of many photon quantum information processing,
is the heralded Hong-Ou-Mandel visibility,
which quantifies the interference of photons heralded from different sources. 
This quantity,
which simultaneously incorporates
source indistinguishability, purity and absence of multi-photon noise,
determines the stochastic noise
in photonic quantum computing architectures ~\cite{rudolph2017optimistic,SparrowNoise},
and the computational complexity achievable in photonic sampling algorithms~\cite{renema2018}.
We implemented heralded HOM experiments
by operating our two-source device in the four-photon regime,
as shown in Fig.~\href{\ref{Fig_HOM}}{\ref{Fig_HOM}a}. 
The circuit is configured such that
idler photons from both sources are directly out-coupled
to detectors to herald the signal photons,
which are interfered in the MZI
(see inset of Fig.~\href{\ref{Fig_HOM}}{\ref{Fig_HOM}b}). 
The heralded HOM fringe is measured
by scanning the phase $\theta_1$
inside the MZI and collecting 4-photon events~\cite{Faruque18,adcock2019}. 
The measured on-chip heralded HOM fringe
is shown in
Fig.~\href{\ref{Fig_HOM}}{\ref{Fig_HOM}b}.
The raw-data visibility (no multi-photon noise correction) is $96(2)\%$. 


Our results have a significant impact on
the prospects of quantum information processing in integrated  photonics.
Photon sources from
previous state-of-the-art integrated photonic devices
demonstrated an on-chip heralded quantum inference raw visibility
of $82\%$~\cite{adcock2019}
which upper-bounds any potential quantum sampling experiment
to a computational complexity
equivalent to $31$-photon interference
(considering error bounds of $10\%$~\cite{renema2018}).
Our results lift this bound to a computational complexity
equivalent to $>150$ photon interference,
deep in the regime of quantum computational supremacy~\cite{Neville2017}.

Furthermore,
in the context of digital quantum computing,
our results make a significant leap toward
the $\gtrsim 99.9\%$ heralded HOM visibility
required to construct lattices of physical qubits
with error rates below $1\%$
using current fault-tolerance photonic architectures~\cite{Gimeno-Segovia2015,SparrowNoise}.
Our analysis
(see Supplementary Information~3)
suggests that heralded HOM visibilities of $99.9\%$ could be
achievable with minor modifications to our design;
for example
by using an improved quality pump laser
and by using semiconductor fabrication processes
with approximately
$4$~nm uniformity~\cite{selvaraja2009subnanometer,lu2017performance}.
Our results represent the near removal of
a critical set of physical errors that
had limited the scaling of
photonic quantum information processing.

\section*{Methods}

{\small

\noindent\textbf{Device fabrication.}
The silicon devices used were fabricated using CMOS-compatible UV-lithography processes in a commercial multi-project wafer run by the Advanced Micro Foundry (AMF) in Singapore. Waveguides are etched in a 220~nm silicon layer atop a 2~$\mu$m buried oxide, and an oxide top cladding of 3 $\mu\text{m}$. The thermo-optic phase-shifters to reconfigure the integrated circuits are formed by TiN heaters positioned 2~$\mu$m above the waveguide layer. \\ \ \\

\noindent\textbf{Inter-modal four-wave mixing in silicon waveguides.}
Inter-modal spontaneous four-wave mixing is performed by propagating the pump, signal and idler waves on different waveguide modes. The spectral properties of the perfect phase matching depend on the group velocity dispersion of the different modes employed in the process, which can be tuned by engineering the waveguide geometry.
In our experiment, we 
operate
the pump on the TM0 and TM1 modes, and the signal and idler on the TM1 and TM0 respectively.
With these modes, phase matching of the SFWM process is enabled in the 1500 - 1600 nm spectral window using standard silicon-on-insulator waveguides with a geometry of 2~$\mu$m $\times$ 0.22~$\mu$m, which is used in our source design. 
\\ \ \\

\noindent\textbf{Pump-delayed generation.}
When a delay is applied between non-degenerate pump pulses, the effective interaction length depends on the value of such delay. This is due to the walk-off, which limits the nonlinear length. The best scenario is when the overtaking process of the faster pulse over the slower one occurs completely within the waveguide, thus maximising  the interaction length and the generation efficiency. In this case the delay is such that the maximum spatial overlap between the pump pulses occurs in the middle of the waveguide. With different delays, the nonlinear medium is not optimally exploited, resulting in reduced generation efficiency. The delay used to optimise the generation efficiency corresponds to the delay for maximum spectral purity (details in Supplementary Information~1).
\\ \ \\

\noindent\textbf{Source design.}
The 2~$\mu$m $\times$ 0.22~$\mu$m multi-mode waveguide in the source is designed with a length of 11~mm and an initial temporal delay of $\tau=1.46~\text{ps}$ between the TM0 and TM1 modes. A spiral geometry for the waveguide is used to increase the compactness. Modal cross-talk in the spiral is kept below $-25$~dB extinction by adopting $90^{\circ}$ Euler bends of radius $45~\mu\text{m}$ (see Supplementary Information~1 for more details). The footprint of an individual silicon-on-insulator source with our design is approximately $200 \mu\text{m} \times900 \mu\text{m}$. The TM0--TM1 mode converters used to inject the pump in the MM waveguide and separate the signal and idler photons at the output have $<-30$~dB characterised modal cross-talk, and $>95\%$ conversion efficiency .
\\ \ \\

\noindent\textbf{Integrated circuit.}
The integrated circuit pictured in Fig.~\href{Fig_Indist}{\ref{Fig_Indist}a} (see also Supplementary Fig.~6 for a more detailed schematic) used for the multi-source interference experiments consists of three reconfigurable MZIs (internal phases $\varphi$, $\theta_1$ and $\theta_2$), two phase-shifters ($\phi_1$, $\phi_2$), a broad-band waveguide crosser, and two sources. The circuit used is a two-mode version of the circuits implemented, for example, in Ref.~\cite{wang2018multidimensional}. At the input, the MZI $\varphi$ is configured to split the pump between the two sources: using $\varphi=0$ ($\varphi=\pi$) we operate the sources individually by pumping only source 1 (source 2), while $\varphi=\pi/2$ implements a balanced pump splitting to coherently operate both sources simultaneously. After photons are generated in the sources, the waveguide crosser allows us to route together to signal and idler modes. Arbitrary and reconfigurable two-mode unitary operations are then performed on the signal (idler) modes via the phase $\phi_1$ ($\phi_2$) and the MZI $\theta_1$ ($\theta_2$). Light is coupled in and out of the circuit by means of TM0 focusing grating couplers, which have been individually optimised to maximise their efficiency at the pump, signal and idler wavelengths ($\simeq 6.6$~dB loss per coupler, see Supplementary Information~2). Total insertion losses in the integrated circuit are approximately -14~dB, mostly due to grating couplers. \\ \ \\

\noindent\textbf{Experimental set-up.} Pump pulses at 1550~nm (4.5~nm bandwidth, 800~fs pulse length, 50~MHz repetition rate) from an erbium-doped fibre laser (Pritel) are filtered via a square-shaped, 5 nm bandwidth filter (Semrock) to eliminate spurious tails at the signal and idler wavelengths, and then injected into the device. A fiber polarisation controller (Lambda) is used to ensure injection of TM0 polarised light to maximise the coupling. After the chip, pump rejection is performed via broadband ($>12$~nm bandwidth, much larger than the photon spectra) band-pass filters (Opneti), and photons are finally detected using  superconducting nanowire single-photon detectors with approximately $80\%$ average efficiency (Photon Spot). For the JSI reconstruction, we use tunable filters with adjustable bandwidth (EXFO XTA-50). Analogue voltage drivers (Qontrol Systems, 300 $\mu$V resolution) are used to drive the on-chip phase shifters and reconfigure the integrated circuit.
}

\bibliography{\jobname}

\begin{footnotesize}

\hfill \vspace{-0.1cm}

\noindent\textbf{Acknowledgements.} We thank N. Maraviglia, J. Bulmer, and F. Graffitti for useful discussions, and L. Kling for technical assistance. 
We acknowledge support from the 
Engineering and Physical Sciences Research Council (EPSRC) Hub in Quantum Computing and Simulation (EP/T001062/1), European Commission QUCHIP (H2020-FETPROACT-3-2014:
quantum simulation), and the European Research Council (ERC). LP and SS have been supported by the University of Trento’s strategic initiative Q@TN, European Union's Horizon 2020 research and innovation programme under grant agreement No 820405.
Fellowship support from EPSRC is acknowledged by A.L. (EP/N003470/1).\\

\noindent\textbf{Competing interests.} S.P., M.B., S.S. and L.P declare UK patent application number 2005827.7.\\

\end{footnotesize}

\end{document}